\documentclass[prd,twocolumn,superscriptaddress,showpacs,nofootinbib,preprintnumbers]{revtex4}

\usepackage{amsmath}
\usepackage{amsfonts}
\usepackage{graphicx}
\usepackage{dcolumn}

\def\be{\begin{equation}}
\def\ee{\end{equation}}
\def\ba{\begin{eqnarray}}
\def\ea{\end{eqnarray}}
\def\bs{\begin{subequations}}
\def\es{\end{subequations}}

\newcommand{\rd}{{\rm d}}
\def\d{{\rm d}}

\newcommand{\tk}{\tilde{k}}

\usepackage{color}

\begin{document}

\title{Inflation and dark energy arising from geometrical tachyons}

\author{Sudhakar Panda}
\affiliation{Harish-Chandra Research Institute, Chhatnag Road,
Jhusi, Allahabad-211019, India}
\email{panda@mri.ernet.in}

\author{M.~Sami}
\affiliation{Department of Physics, Jamia Millia,
New Delhi-110025, India}
\email{sami@iucaa.ernet.in}

\author{Shinji Tsujikawa}
\affiliation{Department of Physics, Gunma National College of
Technology, Gunma 371-8530, Japan}
\email{shinji@nat.gunma-ct.ac.jp}

\date{\today}

\begin{abstract}

We study the motion of a BPS $D3$-brane in the $NS5$-brane ring
background. The radion field becomes tachyonic in this geometrical
set up. We investigate the potential of this geometrical tachyon in
the cosmological scenario for inflation as well as dark energy.
We evaluate the spectra of scalar and tensor perturbations generated
during tachyon inflation and show that this model is compatible
with recent observations of Cosmic Microwave Background (CMB)
due to an extra freedom of the number of $NS5$-branes.
It is not possible to explain the origin of both inflation and dark
energy by using a single tachyon field, since the energy density
at the potential minimum is not negligibly small because of the
amplitude of scalar perturbations set by CMB anisotropies.
However geometrical tachyon can account for dark energy when
the number of $NS5$-branes is large, provided that inflation is
realized by another scalar field.

\end{abstract}

\maketitle

\section{Introduction}

The dynamics of time dependent backgrounds in string theory has
been a challenging problem for long time. Recent progress on
tachyon condensation by Sen \cite{senrev} has been very useful
for studying such time-dependent backgrounds. The rolling tachyon
played an important role in studying the decay of unstable
$D$-branes as well as annihilation of brane-antibrane pairs. The
Dirac-Born-Infeld (DBI) action \cite{DBI} was used, as an
effective field theory description, for the dynamics of this
tachyon. It was observed that as the tachyon condenses, the
unstable brane or the brane-antibrane pair can decay to form a
new stable $D$-brane. This observation has given rise to the original
idea of tachyon cosmology with the hope that the open string
tachyon on the unstable brane can be the scalar field driving
inflation \cite{tachinfl,KL} (see also Refs.~\cite{tachyonpapers}
for the application of tachyon in cosmology). This idea was also
generalized to the radion field, in the case of a  brane moving
towards an antibrane and vice versa \cite{BA}. But problems like
incompatibility of slow-roll, too steep potential and reheating
plagued the development of tachyon cosmology, though some of them
could be solved via warped compactification \cite{GST,warpcosm}
of string theory. Inspite of several
attempts to overcome the problems in open string tachyon
cosmology, it seems unlikely that this tachyon field is
responsible for inflation. Instead it is more likely to play a
role as dark matter fluid \cite{sen}.

Recently the DBI action has found a prominent role in the study of
a different time dependent background. It describes the dynamics
of a $D$-brane in the background of $k$ coincident $NS5$-branes
\cite{kutasov} where the $D$-brane is effectively a probe brane
i.e., it probes the background without disturbing it. The reason
behind this phenomena is that while the tension of the
$NS5$-brane goes as $1/g_s^2$ the tension of the $D$-brane goes
as $1/g_s$ where $g_s$ is the string coupling. Thus $NS5$-branes
are much heavier than the $D$-branes in the regime of small
string coupling. Geometrically this means that the $NS5$-branes
form an infinite throat in space-time and the string coupling
increases as we move towards the bottom. Being lighter, the probe
brane is gravitationally pulled towards the $NS5$-branes. Since
the $D$-brane preserves half of the supersymmetry which is
different from the other half preserved by the $NS5$-branes in
Type-II theory, as the probe brane comes nearer to the source
brane the configuration becomes unstable. The radion becomes
tachyonic and is the source for the instability. Kutasov showed
that there is a map between the tachyonic radion field living on
the world volume of the probe brane and the rolling tachyon
associated with a non-BPS $D$-brane and thus the motion of the
probe brane in the throat could be described by the condensation
of the tachyon. The probe brane thus decays into tachyonic matter
with a pressure that falls off exponentially to zero at late
times. Furthermore, it was shown, by compactifying one of the
transverse direction to the source branes, it is possible to
obtain a potential which resembles the potential obtained by the
use of techniques of string field theory. Thus it is believed
that the tachyon has a geometrical origin.

The above situation has been extended to a different configuration
of the source brane in Refs.~\cite{TW1,TW2}. In this
set-up, these authors examined the motion of the probe brane in the
background of a ring of $NS5$-branes instead of coincident branes
and obtained several interesting solutions for the probe brane in
the near horizon (throat) approximation. It is observed that the
radion field becomes tachyonic when the probe brane is confined to
one dimensional motion inside the ring. Following the cosmological
applications of the original tachyon condensation, the
condensation of the geometrical tachyon has also naturally played a
role in cosmology \cite{NS5cos}. In this paper, we shall carry out 
detailed analysis for the role played by the geometrical tachyons 
in cosmology following the analysis of \cite{TWinf}.
We evaluate the spectra of density perturbations generated in 
geometrical tachyon inflation and place strong constraints on 
model parameters by confronting  them with recent observational data.
We also show an interesting possibility to use geometrical tachyon 
as a source for dark energy.

This paper is organized as follows. In the next section we discuss
the basic set-up for geometrical tachyon. In Sec.~III we shall study
inflation based upon geometrical tachyon and evaluate the spectra
of scalar and tensor perturbations. The model parameters are
constrained using latest observational data coming from CMB. In
Sec.~IV we match tachyon potentials around the ring of the
$NS5$-branes and study the possibility of reheating. In Sec.~V we
shall apply our geometrical tachyon scenario for dark energy. We
summarize our results in the final section.

\section{Geometrical tachyon}

We begin with the background fields around $k$ parallel
$NS5$-branes of type II string theory. In this case the metric is
given by \cite{Callan}
\begin{eqnarray}
\label{eq:ring}
\rd s^2 = \eta_{\mu \nu} \rd
x^{\mu} \rd x^{\nu}
+ F(x^n) \rd x^m \rd x^m\,,
\end{eqnarray}
where the dilaton field $\chi$ is defined as $e^{2(\chi-\chi_{0})} =
F (x^n)$ and the three form field strength associated with 
the $NS$ two form potential is $F_{mnp}= -\varepsilon^q_{mnp}
\partial_q \chi$. Here $F(x^n)$ is the harmonic function describing
the position of the fivebranes. For the fivebranes at generic
positions $x_1,....,x_k$, the harmonic function is found to be
\begin{equation}
F = 1 + l_s^2 \sum_{j=1}^k \frac{1}{|x - x_j|^2}\,,
\end{equation}
where $l_s = \sqrt{\alpha'} $ is the string length. The
fivebranes are stretched in the directions $( x^0, x^1,....,
x^5)$ and are localized in the $(x^6,..., x^9)$ directions. For
coincident fivebranes an $SO(4)$ symmetry group of rotations
around the fivebranes is preserved. This gives rise to a throat
geometry. But we are interested in considering the geometry obtained
from the extremal limit of the rotating $NS5$- brane solution as
discussed in \cite{sfet} where the branes are continuously
distributed along a ring of radius $R$, which is oriented in the
$x^6 - x^7$ plane in the transverse space. The above $SO(4)$
symmetry is thus broken. The full form of the harmonic function
in the throat region is then given by
\begin{eqnarray}
F&=&1+ \frac{kl_s^2}{2R\rho \sinh(y)} \frac{\sinh(ky)}
{(\cosh(ky)-\cos(k\theta))} \\
&\simeq& \frac{kl_s^2}{2R\rho \sinh(y)}
\frac{\sinh(ky)}{(\cosh(ky)-\cos(k\theta))}\,,
\label{Har}
\end{eqnarray}
where $\rho, \theta$ parameterize the coordinates in the ring
plane i.e. $x^6 = \rho \cos(\theta), x^7 = \rho \sin(\theta)$ and
the factor $y$ is given by
\begin{equation}
\label{coshy}
\cosh(y) = \frac{R^2+\rho^2}{2R\rho}.
\end{equation}

\begin{figure}
\includegraphics[height=2.5in,width=2.5in]{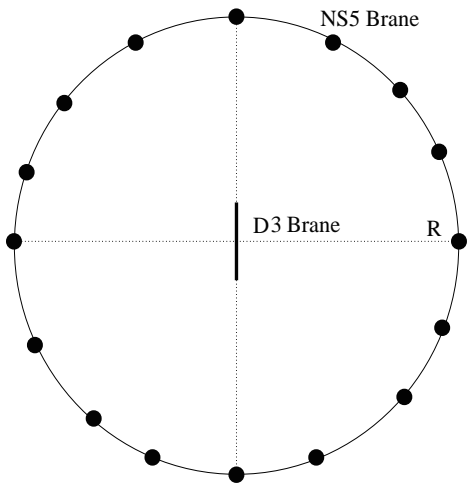}
\caption{Figure showing a $D3$-brane in the static background
of $NS$5-branes located on the circle of radius $R$.}
\label{age}
\end{figure}

We introduce a probe $Dp$-brane at the center of the circle in
the background of $NS$5 branes (see Fig.~\ref{age}). 
As mentioned in the introduction, 
the probe brane will be attracted towards
the ring of $NS5$-branes due to gravitational interaction. The
dynamics of the probe brane can be described by the action 
(see \cite{kutasov,TW1,TW2} for details):
\begin{equation}
S=-\tau_p \int {\rm d}^{p+1} \zeta
\sqrt{F^{-1}-(\dot{\rho}^2+\dot{\sigma}^2)}\,,
\label{action1}
\end{equation}
where $\tau_{p}$ is the brane tension and  $\zeta$ is the
world-volume directions of the probe brane. Here $\sigma$ is the
radial coordinate for $x^8$ and $x^9$ when expressed in polar
coordinates.

The $Dp$ brane will be pulled towards the circumfrance of the
circle and for simplicity we consider the motion in the plane of
the circle ($\sigma=0$). As discussed earlier, the radion field
$\rho$ becomes tachyonic and the action (\ref{action1}) gets
mapped to a well known action describing the dynamics of the
tachyon field living on the world volume of a non-BPS brane in
type II string theory:
\begin{equation}\label{action2}
S= -\int {\rm d}^{p+1}
\zeta V(\phi) \sqrt{1-\dot{\phi}^2},
\end{equation}
where the tachyon map is given by
\begin{equation}
\label{phirho}
\phi(\rho)=\int{\sqrt{F} \d\rho}\,,
\end{equation}
and the tachyon potential is given by
\begin{equation}
\label{tachpo}
V(\phi)=\frac{\tau_p}{\sqrt{F}}\,.
\end{equation}
%

In what follows we shall consider the case of $p=3$.
We look for the motion of $D3$ brane far away from the circumfrance
of the circle where $NS5$-branes are distributed
($\ell \gg {2\pi R}/k $ $-$  distances much larger than brane separation).
In this case the expression of harmonic function simplifies to
\begin{equation}
F=\frac{kl_s^2}
{\sqrt{(R^2+\rho^2 + \sigma^2)^2-4R^2\rho^2}}\,.
\label{eq:ring2}
\end{equation}
The motion of the probe brane in the plane of ring
corresponds to $\sigma=0$. In this case the Harmonic
function yields
\begin{equation}
F=\frac{kl_{s}^2}{R^2-\rho^2}\,.
\end{equation}

{}From Eqs.~(\ref{phirho}) and (\ref{tachpo})
we obtain
\begin{equation}
\phi(\rho) = \sqrt{kl_s^2} \arcsin(\rho/R)\,,
\end{equation}
and
\begin{eqnarray}
\label{eq:cosine}
V(\phi) =  V_0 \cos\left(\frac{\phi}{\sqrt{kl_s^2}}\right)\,,
~~~V_0 = \frac{\tau_3 R}{\sqrt{kl_s^2}}\,.
\end{eqnarray}
Most of the contribution to inflation comes from the top
of the potential corresponding to the motion of $D3$ brane
near the center of the circle.

A throat approximation is used to get the exact expression
above. This demands the following condition
\begin{equation}
\sqrt{k}l_{s} \gg R\,.
\label{throat}
\end{equation}
We note that the reduced Planck mass, $M_{p}=1/\sqrt{8\pi G}$,
is related with the string mass scale, $M_{s}=1/l_s$,
via the dimensional reduction:
\begin{equation}
M_{p}^2=\frac{vM_{s}^2}{g_{s}^2}\,,
\label{Mv}
\end{equation}
where $g_{s}$ is the string coupling parameter and
\begin{equation}
v \equiv \frac{(M_{s} l)^d}{\pi}
=\frac{1}{\pi} \left(\frac{l}{l_{s}}\right)^d\,.
\end{equation}
Here $l$ and $d$ are the radius and the number of
compactified dimensions, respectively.
In order for the validity of the effective theory we require
the condition $l_{s} \ll l$, which translates into
the condition
\begin{equation}
v \gg 1\,.
\end{equation}
%

\section{Inflation}

In a flat Friedmann-Robertson-Walker (FRW) background
with a scale factor $a$, the evolution equations are
given by \cite{tachyonpapers}
\ba
& & H^2
=\frac{1}{3M_p^2}
\frac{V(\phi)}{\sqrt{1-\dot{\phi}^2}}\,, \\
& & \frac{\ddot{\phi}}{1-\dot{\phi}^2}+3H\dot{\phi}
+\frac{V_{\phi}}{V}=0\,,
\ea
where a dot denotes a derivative in terms of
the cosmic time $t$ and $H \equiv \dot{a}/a$ is the Hubble rate.
{}From the above equations we obtain
\begin{equation}
\frac{\ddot{a}}{a}=\frac{V(\phi)}{3M_p^2\sqrt{1-\dot{\phi}^2}}
\left(1-\frac32 \dot{\phi}^2\right)\,.
\end{equation}
Hence the inflationary phase ($\ddot{a}>0$)
corresponds to $\dot{\phi}^2<2/3$.
Since the energy density and the pressure density of tachyon are
given by $\rho=V(\phi)/\sqrt{1-\dot{\phi}^2}$ and
$p=-V(\phi)\sqrt{1-\dot{\phi}^2}$, the equation of state is
\begin{equation}
w_{\phi} \equiv \frac{p}{\rho}=\dot{\phi}^2-1\,.
\label{eqstate}
\end{equation}
\subsection{Background}

By using a slow-roll approximation, $H^2 \simeq V(\phi)/3M_{p}^2$
and $3H\dot{\phi} \simeq -V_{\phi}/V$,
the number of e-foldings, $N={\rm ln}\,a$, is
\ba
\label{efolds}
N&=&\int_{t}^{t_{f}} H {\rm d}t=\int_{\phi_{f}}^\phi
\frac{V^2}{M_{p}^2V_{\phi}} {\rm d}\phi \nonumber \\
&=&
s \left[ \cos x_{f}-\cos x+{\rm ln} \left(
\frac{\tan (x_f/2)}{\tan (x/2)} \right) \right]\,,
\ea
where the subscript ``$f$'' represents the values at the end of
inflation and
\ba
\label{spara}
s \equiv \frac{\tau_{3}R\sqrt{kl_{s}^2}}
{M_p^2},~~
x \equiv \frac{\phi}{\sqrt{kl_s^2}},~~
x_{f} \equiv \frac{\phi_{f}}{\sqrt{kl_s^2}}.
\ea

The slow-roll parameter, $\epsilon \equiv -\dot{H}/H^2$,
is given by
\begin{equation}
\epsilon=\frac{1}{2s}
\frac{\sin^2 x}{\cos^3 x}\,.
\end{equation}
It is convenient to introduce the following quantities:
\ba
y \equiv \cos x\,,~~~
y_{f} \equiv \cos x_{f}\,.
\ea
Then $N$ and $\epsilon$ can be expressed in terms
of $y$:
\ba
\label{efold}
N=s \left[ y_{f}-y+\frac12 {\rm ln}
\frac{(1-y_{f})(1+y)}{(1+y_{f})(1-y)}
\right]\,,
\ea
and
\ba
\epsilon=\frac{1}{2s}
\frac{1-y^2}{y^3}\,.
\ea

The end of inflation is characterized by $\epsilon=1$,
which gives
\ba
y_f \equiv f(s)=
\frac{1}{6s} \left[g(s)+\frac{1}{g(s)}-1\right]\,,
\ea
where
\ba
g(s) \equiv [54s^2-1+6s \sqrt{3(27s^2-1)}]^{1/3} \,.
\ea
{}From Eq.~(\ref{efold}) we have
\ba
\label{difeq}
{\rm ln}\,\frac{1+y}{1-y}-2y
=\frac{2N}{s}-2f(s)-{\rm ln}\,
\frac{1-f(s)}{1+f(s)}\,.
\ea

\begin{figure}
\includegraphics[height=2.9in,width=3.2in]{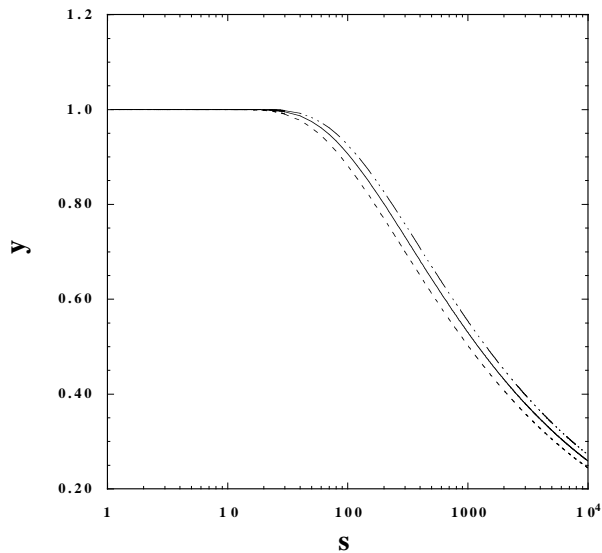}
\caption{The function $y=\cos(\phi/\sqrt{kl_{s}^2})$
versus the parameter $s$ given by Eq.~(\ref{spara}).
Each case corresponds to the number of e-foldings:
$N=70, 60, 50$
from top to bottom.}
\label{ysfig}
\end{figure}

One can not obtain analytic expression for $y$
in terms of $s$ and $N$.
In order to find $y$ as a function of $s$ for a fixed $N$,
it is convenient to take the derivative of Eq.~(\ref{difeq}):
\ba
\frac{\rd y}{\rd s}=
\frac{1-y^2}{y^2} \left[ \frac{f'(s)f^2(s)}{1-f^2(s)}
-\frac{N}{s^2} \right]\,.
\ea
{}From Eq.~(\ref{difeq}) one can find a value of $y$
for a given $s$.
For example we have $y=0.99505941$ for $s=30$
with $N=60$. Then we get $y(s)$ by numerically solving
Eq.~(\ref{difeq}) for a fixed $N$.
In Fig.~\ref{ysfig} we show $y$ versus $s$ for
$N=50, 60, 70$.
The function $y$ gets smaller for larger $s$,
which means that inflation can be realized even if
the field $\phi$ is not very close to the top of the potential.

%
\subsection{Perturbations}

Let us consider a general perturbed metric about the flat FRW
background \cite{MFB}:
\begin{eqnarray}
\rd s^2 &=& -(1+2A)\rd t^2 + 2a(t)B_{,i} \rd x^i \rd t  \nonumber\\
& & +a^2(t)[(1-2\psi) \delta_{ij}+2E_{,i,j}+2h_{ij}] 
\rd x^i \rd x^j.
\nonumber \\
\label{pmetric}
\end{eqnarray}
Here $A$, $B$, $\psi$, and $E$ correspond to the scalar-type
metric perturbations, whereas $h_{ij}$ characterizes the
transverse-traceless tensor-type perturbation. 
We introduce comoving curvature perturbations, ${\cal R}$,
defined by
\begin{eqnarray}
{\cal R} \equiv \psi+\frac{H}{\dot{\phi}}\delta \phi\,,
\label{calR}
\end{eqnarray}
where $\delta \phi$ is the perturbation of the field $\phi$.

By using a slow-roll approximation the power spectrum
of curvature perturbations is estimated to be \cite{HN}
\begin{eqnarray}
{\cal P}_{\rm S}=\left(\frac{H^2}{2\pi\dot{\phi}} \right)^2
\frac{1}{Z_{\rm S}}\,,
\label{powersca}
\end{eqnarray}
where $Z_{\rm S}=V(1-\dot{\phi}^2)^{-3/2} \simeq V$.
The spectral index of scalar perturbations is defined by 
$n_{\rm S}-1 \equiv {\rm d} {\rm ln}
{\cal P}_{\rm S}/{\rm d} {\rm ln}\,\tilde{k}$, where
$\tilde{k}$ is a comoving wavenumber.
Then we obtain 
\begin{eqnarray}
n_{\rm S}-1=2(2\epsilon_1-\epsilon_2-\epsilon_3)\,,
\label{indexsca}
\end{eqnarray}
where
\begin{eqnarray}
\epsilon_1 \equiv \frac{\dot{H}}{H^2}\,,~~~ \epsilon_2 \equiv
\frac{\ddot{\phi}}{H\dot{\phi}}\,,~~~ \epsilon_3 \equiv
\frac{\dot{Z}_{\rm S}}{2HZ_{\rm S}}\,. 
\label{slow}
\end{eqnarray}

The amplitude of tensor perturbations 
is given by \cite{HN}
\begin{eqnarray}
{\cal P}_{\rm T}=8\left(\frac{H}{2\pi}\right)^2\,.
\label{powerten}
\end{eqnarray}
The spectral index, $n_{\rm T} \equiv {\rm d} {\rm ln}
{\cal P}_{\rm T}/{\rm d} {\rm ln}\,\tilde{k}$, is
\begin{eqnarray}
n_{\rm T}=2\epsilon_1\,. \label{indexten}
\end{eqnarray}
We obtain the tensor-to-scalar ratio, as
\begin{eqnarray}
r \equiv \frac{{\cal P}_{\rm T}}{{\cal P}_{\rm S}}
=8\frac{\dot{\phi}^2}{H^2}Z_{\rm S}\,. \label{ratio}
\end{eqnarray}

Using a slow-roll analysis the above quantities can be expressed
by the slopes of the potential:
\begin{eqnarray}
\label{nS}
 {\cal P}_{\rm S} &=& \frac{1}{12\pi^2M_p^6}
\left(\frac{V^2}{V_{,\phi}}\right)^2,\\
\label{nT}
n_{\rm S}-1 &=& -4\frac{M_p^2V_{,\phi}^2}{V^3}
+2\frac{M_{p}^2V_{,\phi \phi}}{V^2},\\
\label{rati}
n_{\rm T} &=& -\frac{V_{,\phi}^2M_{p}^2}{V^3},\\
\label{indexsca2}
r &=& 8\frac{V_{,\phi}^2M_{p}^2}{V^3}\,,
\end{eqnarray}
where we reproduced the Planck mass for a later convenience.
Equations (\ref{rati}) and (\ref{indexsca2})  show 
that the same consistency relation, $r=-8n_{\rm T}$,
holds as in the Einstein gravity \cite{SV,GST}.

\subsection{Constraints on model parameters}

For the potential (\ref{eq:cosine}) the amplitude of scalar
perturbations is given by
\begin{eqnarray}
{\cal P}_{\rm S}&=&\frac{kl_s^2V_{0}^2}{12\pi^2 M_p^6}
\left(\frac{\cos^2 x}
{\sin x} \right)^2 \nonumber \\
&=& \frac{s^2}{12\pi^2k(l_sM_p)^2}
\frac{y^4}{1-y^2}\,.
\end{eqnarray}
The COBE normalization corresponds to
$P_{\rm S} \simeq 2 \times 10^{-9}$ for the mode
which crossed the Hubble radius 60
e-foldings before the end of inflation.
Then this gives the following constraint:
\begin{eqnarray}
k(l_{s}M_{p})^2 \simeq
\frac{10^9}{12\pi^2}
\frac{s^2y^4}{1-y^2}\,.
\label{kcon}
\end{eqnarray}

\begin{figure}
\includegraphics[height=2.9in,width=3.2in]{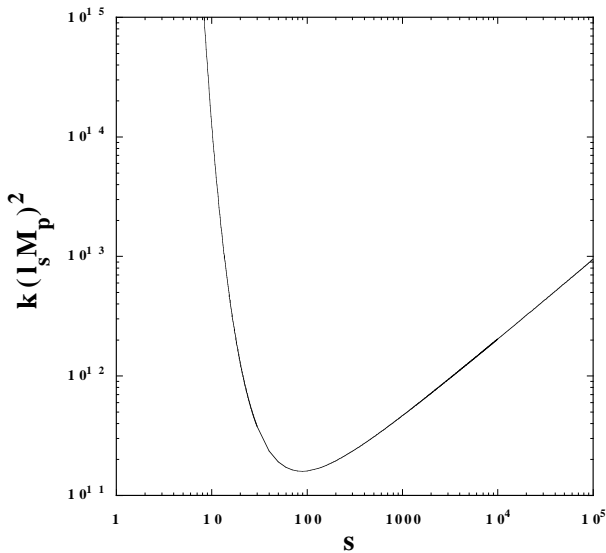}
\caption{$k(l_{s}M_{p})^2$ versus $s$ coming from
the COBE normalization for the mode
which crossed the Hubble radius $60$ e-foldings
before the end of inflation.}
\label{kLMfig}
\end{figure}

In Fig.~\ref{kLMfig} we plot $k(l_{s}M_{p})^2$ in terms of
the function $s$ for $N=60$.  This quantity has a minimum
around $s=100$, which gives a constraint
\begin{eqnarray}
\label{kcon1}
k(l_{s}/l_{p})^2 \gtrsim 10^{11}\,,
\end{eqnarray}
where $l_{p}=1/M_{p}$ is the Planck length.

Let us consider the limiting case: $s \gg 1$.
For a fixed value of $N$, the r.h.s. of
Eq.~(\ref{difeq}) approaches zero for $s \to \infty$.
Comparing this with the l.h.s. of
Eq.~(\ref{difeq}), we find that $y \to 0$ for
$s \to \infty$.
By carrying out a Taylor expansion around $y=0$
and taking note that $f(s)$ behaves as
$f(s) \simeq (2s)^{-1/3}$, we get the relation
$y^3 \simeq (3N+1/2)/s$.
Then we find
\begin{eqnarray}
k(l_{s}M_{p})^2 \simeq
\frac{10^9}{24\pi^2}
(3N+1)^{4/3}s^{2/3}\,.
\label{klimit}
\end{eqnarray}
This means that $k(l_{s}M_{p})^2 \to \infty$
in the limit $s \to \infty$.

\begin{figure}
\includegraphics[height=2.9in,width=3.2in]{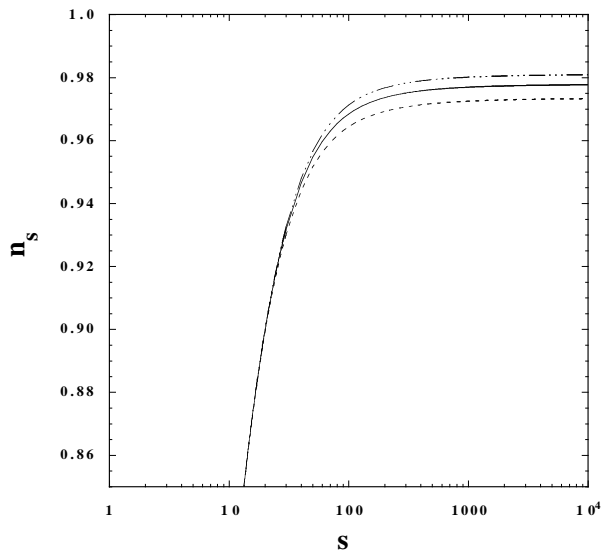}
\caption{The spectral index of scalar metric perturbations
($n_{\rm S}$) as a function of $s$.
Each case corresponds to the number of e-foldings:
$N=70, 60, 50$
from top to bottom.}
\label{nsfig}
\end{figure}

\begin{figure}
\includegraphics[height=2.9in,width=3.2in]{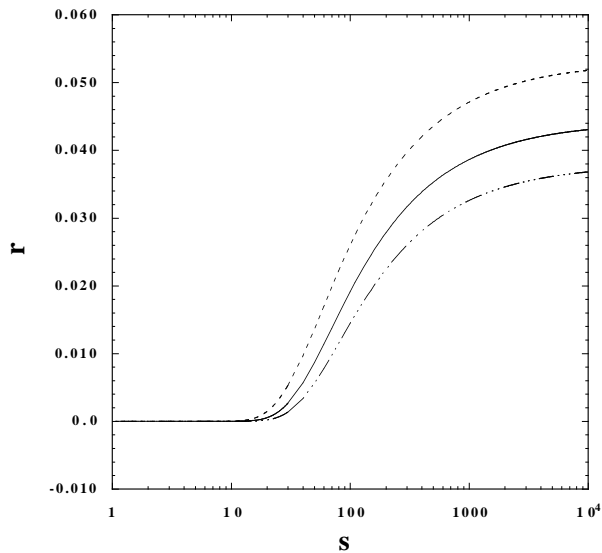}
\caption{The tensor-to-scalar ratio $r$ as a function of $s$.
Each case corresponds to the number of e-foldings: $N=50, 60, 70$
from top to bottom.}
\label{rafig}
\end{figure}

{}From Eqs.~(\ref{nT}), (\ref{rati}) and (\ref{indexsca2})
we obtain
\begin{eqnarray}
\label{n}
& & n_{\rm S}-1=-\frac{2}{s}
\frac{2-y^2}{y^3}\,,\\
& & n_{\rm T}=
-\frac{1}{s} \frac{1-y^2}{y^3}\,, \\
\label{rat}
& & r=\frac{8}{s} \frac{1-y^2}{y^3}\,.
\end{eqnarray}
For a fixed value of $N$ one can know $y$ in
terms of the function of $s$.
In Figs.~\ref{nsfig} and \ref{rafig} we plot $n_{\rm s}$
and $r$ versus $s$ for several numbers of e-foldings.
The spectrum of the scalar perturbations is red-tilted
($n_{\rm S}<1$).
We find that $n_{\rm S}$ gets larger with the increase of
$s$. Recent observations show that $n_{\rm S}$ ranges
in the region $n_{\rm S}>0.93$
at 2$\sigma$ level \cite{obcon,obcon2}.
If the cosmologically relevant scales correspond to
$N=60$, the parameter $s$ is constrained to be $s>29.3$.
This value is not strongly affected by the change of
the number of e-foldings, since $s>30.0$ for $N=50$
and $s>28.9$ for $N=70$.
Hence we have the following constraint:
\begin{eqnarray}
\label{con1}
s \gtrsim 30\,.
\end{eqnarray}

Since $y$ behaves as $y^3 \simeq (3N+1/2)/s$
in the limit $s \to \infty$,
Eqs.~(\ref{n}) and (\ref{rat}) yield
\begin{eqnarray}
\label{limit}
n_{\rm S}=1-\frac{8}{6N+1},~~
r=\frac{16}{6N+1},~~~(s \to \infty).
\end{eqnarray}
This means that asymptotic values of $n_{\rm S}$
and $R$ are constants.
When $N=60$, for example, we have
$n_{\rm S}=0.978$ and $r=0.044$.
We checked that $n_{\rm S}$
and $r$ actually approach these values
in numerical calculations in Figs.~\ref{nsfig} and \ref{rafig}.

The tensor-to-scalar ratio is constrained to be
$r<0.36$ at the 2$\sigma$ level from
recent observations \cite{obcon2}.
The ratio $r$ obtained in Eq.~(\ref{limit}) corresponds to
the maximum value, since $r$ is a growing function
with respect to $s$ as illustrated in Fig.~\ref{rafig}.
Since $r_{\rm max}=0.053$ even for $N=50$,
the tensor-to-scalar ratio is small enough to satisfy
observational contour bounds for {\it any} values
of $s$. Hence tensor perturbations do not provide
constraints on model parameters from
current observations.

We note that the Hubble rate during inflation is
approximately given by
$H \simeq [\tau_3 R/(3M_p^2\sqrt{kl_{s}^2})]^{1/2}$.
Meanwhile the effective mass squared of the tachyon is 
$M^2 \equiv V_{\phi \phi}/V=
-1/kl_{s}^2$.
Hence under the condition (\ref{con1}) we find that
\begin{eqnarray}
|M| \lesssim H/\sqrt{10}\,.
\end{eqnarray}
Then the tachyon mass is smaller than the Hubble rate, which is
required to give rise to inflation.

Taking note that the brane tension $\tau_{3}$ is related with a
string
coupling $g_{s}$ via $\tau_{3}=M_{s}^4/(2\pi)^3g_{s}$,
the condition (\ref{con1}) gives
\begin{eqnarray}
g_{s} =s \frac{(2\pi)^3}{\sqrt{k}RM_{s}}
\gtrsim 30 \frac{(2\pi)^3}{\sqrt{k}RM_{s}}\,,
\label{scoupling}
\end{eqnarray}
where we also used Eq.~(\ref{Mv}).
In order for the effective theory of tachyon to be valid,
we require that we are in a weak coupling regime
($g_{s} \ll 1$). Then one obtains the
constraint: $\sqrt{k}RM_{s} \gg 10^4$.
Combining this with the throat condition (\ref{throat}),
we find
\begin{eqnarray}
\sqrt{k}l_{s}
\gg R \gg \frac{10^4}{\sqrt{k}}l_{s}\,.
\end{eqnarray}
This shows that the number of $NS5$ branes at least satisfies
the condition
\begin{eqnarray}
k \gg 10^4\,.
\label{k}
\end{eqnarray}

If we fix the parameter $s$, we can know the value $k(l_s/l_p)^2$
from the information of COBE normalization (see Fig.~\ref{kLMfig}).
For example one has $k(l_s/l_p)^2 \simeq 10^{11}$ for $s=100$.
In this case the string length scale is constrained to be
$l_s \ll 3 \times 10^3 l_{p}$ (or $M_s \gg 3 \times 10^{-4} M_{p}$)
from Eq.~(\ref{k}).
Since $l_{s}$ is expected to be larger than $l_{p}$, we also find
that $k \lesssim 10^{11}$ from the condition
$k(l_s/l_p)^2 \simeq 10^{11}$.
Hence the number of $NS5$-branes is constrained to be
$10^4 \ll k \lesssim 10^{11}$ for $s=100$.

\begin{figure}
\includegraphics[height=2.9in,width=3.2in]{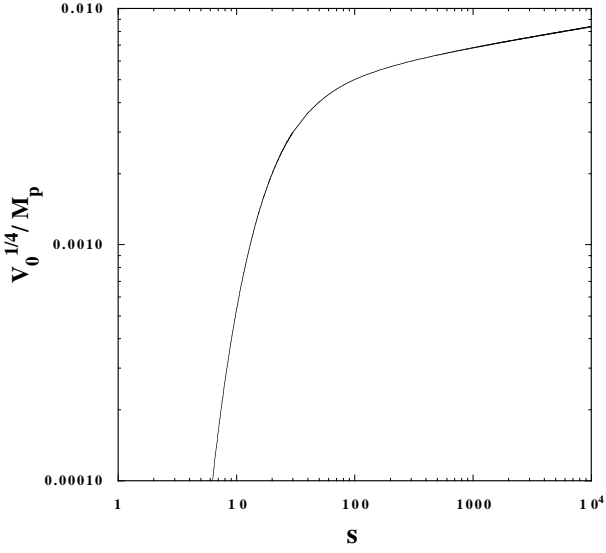}
\caption{The energy scale of inflation, $V_{0}^{1/4}$,
as a function of $s$ for the $60$ e-foldings
before the end of inflation.}
\label{escale}
\end{figure}

Let us consider the energy scale of geometrical tachyon inflation.
{}From Eqs.~(\ref{eq:cosine}) and (\ref{spara}) we find
\ba
V_0=\frac{s}{k(l_sM_p)^2}M_p^4\,.
\ea
In Fig.~\ref{escale} we plot the energy scale $V_0^{1/4}$
as a function of $s$ for the $N=60$ e-foldings before
the end of inflation.
For example one has $V_{0} \simeq 7.9 \times
10^{-11}M_p^4$ for $s=30$ and
$V_{0} \simeq 6.2 \times
10^{-10}M_p^4$ for $s=100$.

In the limit $s \gg 1$ we have
\ba
V_0=\frac{24\pi^2(3N+1)^{-4/3}}{10^9}
s^{1/3} M_p^4\,.
\ea
For the e-foldings $N=60$ this is estimated as
$V_0 \simeq 10^{-9}s^{1/3}M_p^4$.
Hence $V_0$ gets larger with the increase of $s$
as shown in Fig.~\ref{escale}.
{}From the requirement $V_0 \lesssim M_p^4$
one finds that $s$ is constrained to be
$s \lesssim 10^{27}$.
Then by using Eq.~(\ref{klimit}) with e-foldings $N=60$
we find the bound for the quantity $k(l_sM_p)^2$, as
$k(l_sM_p)^2 \lesssim 10^{27}$.
Hence together with Eq.~(\ref{kcon1}), we obtain
\ba
\label{knewcon}
10^{11} \lesssim k(l_s/l_p)^2 \lesssim 10^{27}\,.
\ea

One can not take the limit $k \to \infty$ when we use 
geometrical tachyon for inflation
because of the condition of COBE normalization.
In the context of dark energy, however, we do not have
the restriction coming from the perturbations.
Actually the amplitude of density perturbations should be 
negligibly small in the latter case, 
which means that $10^9$ factor in
Eq.~(\ref{klimit}) is replaced for a very large value.
Then $k(l_sM_p)^2$ can be very large, thereby giving
very small $V_0$ compared to the value obtained 
in inflation. In Sec.~V we shall study the case in which 
geometrical tachyon is used for dark energy.

\section{Reheating}

In this section we shall study the dynamics of reheating 
for the geometrical tachyon.
The tachyon potential has a minimum with a non-vanishing 
energy density at the ring of $NS5$-brane, as we see below.
The field oscillates around the potential minimum 
with a decreasing amplitude by cosmic expansion.

\subsection{Matching tachyon potentials}

The tachyon potential (\ref{eq:cosine}) can be used
inside the ring of the $NS5$-branes ($\rho<R$).
Outside the ring ($\rho>R$) the harmonic function is
given by
\ba
F=\frac{kl_s^2}{\rho^2-R^2}\,.
\ea
{}From Eq.~(\ref{phirho}) we find
\ba
\phi=
\sqrt{kl_s^2}\,{\rm ln}\, \left(\frac{\rho}{R}
+\sqrt{\frac{\rho^2}{R^2}-1}\right)
+\phi_0\,,
\ea
where $\phi_0$ is an integration constant.
Then the tachyon potential is
\ba
\label{po2}
V(\phi)=V_{0}
\sin {\rm h} \left(
\frac{\phi-\phi_0}{\sqrt{kl_s^2}}
\right)\,.
\ea
Here $\phi_0$ corresponds to $\phi_0=(\pi/2)\sqrt{kl_s^2}$.

The two potentials (\ref{eq:cosine}) and (\ref{po2})
can be connected at $\rho=R$ by studying the case
in which $NS5$-branes are not smeared out around
the ring \cite{TWinf}.
Let us consider an expansion around $\rho=R$, i.e.,
$\rho=R+\xi$ with $|\xi| \ll R$.
By substituting this for Eq.~(\ref{coshy}), we find the
following relation
\be
y={\rm ln} (z+1)\,,~~~z \equiv \xi/R\,.
\ee
Then the Harmonic function (\ref{Har}) is
approximately given by
\ba
F &\simeq& \frac{k^2l_s^2}{2R^2(1-\cos (k\theta))}
\nonumber \\
& & \times
\left[1-z+\left\{\frac{5}{6}-
\frac{k^2(2+\cos(k\theta))}{6(1-\cos(k\theta))}
\right\}z^2\right].
\ea

{}From Eq.~(\ref{phirho}) and (\ref{tachpo})
we obtain
\ba
\hspace*{-1.0em}
V&=& \frac{\tau_3 \sqrt{2R^2(1-\cos (k\theta))}}
{kl_s} \nonumber \\
\hspace*{-1.0em}
& & \times
\left[1+\frac{z}{2}+
\left\{\frac{k^2(2+\cos(k\theta))}{12(1-\cos(k\theta))}
-\frac{1}{24}
\right\}z^2\right].
\ea
and
\ba
\hspace*{-1.5em}
\phi &=& \phi_0+ \frac{kl_s}{\sqrt{2(1-\cos(k\theta))}}
\nonumber \\
\hspace*{-1.0em}
& & \times
\left[x-\frac{z^2}{2}+
\left\{\frac{5}{6}-
\frac{k^2(2+\cos(k\theta))}{6(1-\cos(k\theta))}
\right\}\frac{z^3}{3} \right].
\label{ph}
\ea
By taking the first term in Eq.~(\ref{ph})
we find
\ba
\label{po3}
\hspace*{-2.0em}
V(\phi)&\simeq& V_1
\Biggl[1+\frac{\sqrt{2(1-\cos(k \theta))}}
{2kl_s}(\phi-\phi_0) \nonumber \\
\hspace*{-2.0em}
&& +\left( \frac{2+\cos(k \theta)}{6l_s^2}
-\frac{1-\cos (k\theta)}{12k^2l_s^2}\right)
(\phi-\phi_0)^2 \biggr],
\ea
where
\ba
\label{V1}
V_1=\frac{\tau_3 R\sqrt{2
(1-\cos(k\theta))}}{kl_s}\,.
\ea

One can connect  two potentials
(\ref{eq:cosine}) and (\ref{po3})
at $\phi=\phi_0-\epsilon_1 \sqrt{kl_s^2}$.
By matching the potentials with the continuity condition of
$V(\phi)$ together with that of $\rd V(\phi)/\rd \phi$,
we obtain
\ba
& &
\epsilon_1 = \frac{3}{\sqrt{k} (2+\cos(k\theta))
\sqrt{2(1-\cos(k\theta))}}\,, \\
& &
\cos (k\theta) = \frac{-1+\sqrt{6}}{2}\,.
\label{coskt}
\ea
In deriving these values we used the condition
(\ref{k}), under which the terms including $k$
in the denominator in Eq.~(\ref{po3})
are neglected.
Then the tachyon potential around $\rho=R$
is approximately given by
\ba
\label{po3d}
V(\phi) \simeq V_{1} \left[1+\frac{2+\cos(k\theta)}
{6l_{s}^2}(\phi-\phi_{0})^2\right]\,.
\ea

Two potentials (\ref{po2}) and (\ref{po3})
can be also matched at $\phi=\phi_0+\epsilon_2
\sqrt{kl_s^2}$.
One easily finds that $\epsilon_{2}=\epsilon_{1}$
together with $\cos (k\theta) = (-1+\sqrt{6})/2$
under the condition of Eq.~(\ref{k}).

\subsection{The dynamics of reheating}

The energy density at the potential minimum
is given by Eq.~(\ref{V1}).
The ratio of $V_{1}$ and $V_{0}$ is
\ba
\label{V1V0}
\frac{V_{1}}{V_{0}}=
\sqrt{\frac{2(1-\cos(k \theta)}{k}}
 \sim \frac{1}{\sqrt{k}}\,,
\ea
where we used Eq.~(\ref{coskt}).
Hence the energy density $V_1$ is suppressed by the
factor $1/\sqrt{k}$ compared to the energy scale of
inflation.

\begin{figure}
\includegraphics[height=2.9in,width=3.2in]{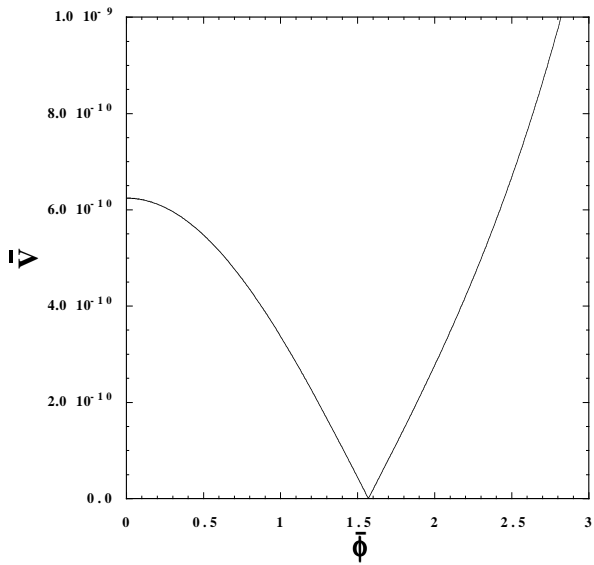}
\caption{The potential of the geometrical tachyon which is
matched around the ring of the $NS5$ branes.
$V$ and $\phi$ are normalized as
$\bar{V}=V/M_p^4$ and $\bar{\phi}=\phi/\sqrt{kl_{s}^2}$.
This plot corresponds to $s=10^2$ and $k=10^6$.
}
\label{potential}
\end{figure}

\begin{figure}
\includegraphics[height=2.7in,width=3.0in]{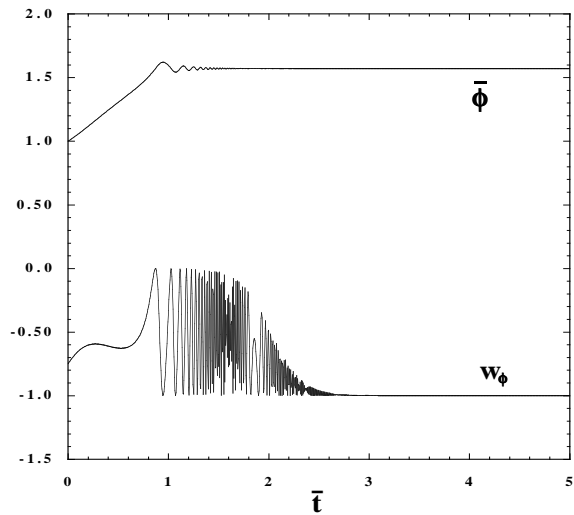}
\caption{The evolution of the tachyon field $\phi$
and its equation of state $w_{\phi}$ for the same
model parameters given in Fig.~\ref{potential}.
$\phi$ and $t$ are normalized as
$\bar{\phi}=\phi/\sqrt{kl_{s}^2}$ and $\bar{t}=t/\sqrt{kl_{s}^2}$.
We choose initial conditions $\phi_{i}=\sqrt{kl_{s}^2}$
and $\dot{\phi}_{i}=0.5$. The field is eventually
trapped at the potential minimum
($\phi=(\pi/2)\sqrt{kl_{s}^2}$) after oscillations.
}
\label{phiw}
\end{figure}

In Fig.~\ref{potential} we plot the tachyon potential
for $s=10^2$ and $k=10^6$, which is obtained by using
the matching condition given in the previous
subsection. We recall that the energy scale $V_0$
is determined by COBE normalization, e.g.,
$V_{0}=6.2 \times 10^{-10}M_p^4$ for $s=10^2$.
When $k=10^6$ the energy scale $V_{1}$
at the potential minimum is $10^{-3}$ times smaller
than $V_{0}$.
In Fig.~\ref{phiw} we plot the evolution of the tachyon field
$\phi$ together with its equation of state $w_{\phi}$
for $s=10^2$ and $k=10^6$.
The field oscillates around the potential minimum
and eventually settles down at
$\phi=(\pi/2)\sqrt{kl_s^2}$.
The equation of state of the tachyon
approaches the one of cosmological constant ($w_{\phi}=-1$),
as is illustrated in Fig.~\ref{phiw}.

Unless the energy scale $V_{1}$ is negligibly small
compared to $V_{0}$, this energy density
comes to dominate the universe in radiation or matter dominant
era, which disturbs the thermal history of the universe.
We note that the number of branes is constrained to
be $k \lesssim 10^{27}(M_{s}/M_{p})^2$
from Eq.~(\ref{knewcon}).
For example one has $V_{1} \simeq 3 \times 10^{-14}V_{0}$
for $M_{s}=M_{p}$. Since $V_{0}$ is larger than of order
$10^{-10}M_p^4$ from Fig.~\ref{escale}, we find
$V_1 \gtrsim 10^{-24}M_{p}^4$.
This energy scale is still too high to explain the late-time
acceleration of the universe which is observed today.
This situation does not change much even if
the string energy scale is smaller than $M_p$
to realize larger $k$
(We note that it is not natural to consider the case in which
$M_s$ is far below $M_{p}$).

In the context of open-string tachyon it was shown in
Ref.~\cite{FKS} that there is a violent instability of
tachyon fluctuations for the potential with a minimum, e.g.,
$V(\phi)=(1/2)m^2(\phi-\phi_0)^2$.
In a flat FRW background the each Fourier mode of
the perturbation in $\phi$ satisfies the following equation
of motion \cite{FKS,GST}:
\begin{eqnarray}
\label{phik}
\frac{\ddot{\delta\phi}_{\tk}}{1-\dot{\phi}^2}
&+&
\left[3H+\frac{2\dot{\phi}\ddot{\phi}}{(1-\dot{\phi}^2)^2}
\right]\dot{\delta \phi}_{\tk} \nonumber \\
&+&\left[\frac{\tk^2}{a^2}+({\rm ln}
V),{}_{\phi\phi}\right]\delta\phi_{\tk}=0\,,
\end{eqnarray}
where $\tilde{k}$ is a comoving wavenumber.
One easily finds that the $(\log V),{}_{\phi\phi}$ term exhibits
a divergence at  $\phi=\phi_{0}$
for the potential $V(\phi)=(1/2)m^2(\phi-\phi_0)^2$.

One may worry that there is a similar instability
for geometrical tachyon around the potential minimum, but
this is not the case.
For the potential (\ref{po3d}) we find
\begin{eqnarray}
({\rm ln} V),{}_{\phi\phi}=
\frac{2c[1-c(\phi-\phi_0)]^2}
{[1+c(\phi-\phi_0)^2]^2}\,,
\end{eqnarray}
where $c=[2+\cos(k \theta)]/6l_s^2$.
This means that there is no divergence in the
denominator at $\phi=\phi_0$, thereby showing the
absence of a violent instability around
the potential minimum.

In spite of the fact that the violent growth of perturbations can
be avoided in our model, we need to find out a way to
avoid that the tachyon energy density overdominates the
universe after inflation. This may be possible if there exists
a negative cosmological constant which almost cancels
with the energy density $V_{1}$
as in Ref.~\cite{GST}.

\section{Dark energy}

The discussion in sections III and IV shows that the
energy scale at the potential minimum is too large
not to disturb the thermal history of the universe
{\it if} we use the information of density perturbations
generated during inflation.
Alternatively let us consider a scenario in which inflation
is realized by a scalar field other than the geometrical
tachyon. In this case we are free from the
constraint (\ref{kcon})  coming from the COBE
normalization. We recall that the COBE
normalization (\ref{kcon}) is used
in order to derive the upper limit of $k$
in Eq.~(\ref{knewcon}).
This upper bound for $k$ is the reason why
the energy density at the potential minimum
overdominates the universe during radiation or
matter dominant era.
If the geometrical tachyon is not responsible for inflation,
we do not have such an upper limit for $k$.
This shows that it may be possible to explain the origin of
dark energy if the value of $k$ is very large.

In order to explain the origin of dark energy
we require the condition
\begin{eqnarray}
V_1 \simeq 10^{-123}M_{p}^4\,.
\label{Lam}
\end{eqnarray}
By using Eq.~(\ref{V1}) together with the matching condition
(\ref{coskt}), we find that $V_1 \simeq \tau_{3}R/(kl_{s})$.
Hence the number of $NS5$-branes is constrained to be
\begin{eqnarray}
k \simeq 10^{123} \frac{\tau_{3}}{M_{p}^4}
\frac{R}{l_s}\,.
\label{dark}
\end{eqnarray}
When $\tau_{3}=10^{-10}M_{p}^4$ (around GUT scale) and $R=10^2
l_{s}$, for example, one has $k=10^{115}$. Since $R$ is at least
greater than $l_{s}$ for the validity of the effective string
theory, we find that $k \gtrsim 10^{123} \tau_{3}/M_p^4$. Hence
we require very large values of $k$ unless the brane tension
$\tau_{3}$ is very much smaller than the Planck scale. We note
that such large values of $k$ automatically satisfy the throat
condition (\ref{throat}).

The energy scale $V_{0}$ is $\sqrt{k}$ times bigger than
$V_{1}$ by Eq.~(\ref{V1V0}).
When $k=10^{115}$ mentioned above, this corresponds to
$V_{0} \simeq 10^{-66}M_{p}^4$ (around TeV scale).
Therefore the tachyon has a considerable amount of energy
when it begins to roll down from the the top of the hill.
However if the tachyon settles down at the potential minimum
in the early universe (before the TeV scale), the energy
density $V_{1}$ of the tachyon does not affect the thermal history of
the universe until it comes out around present epoch
as dark energy.

\section{Conclusions}

In this paper we have studied geometrical tachyon based
upon the movement of a BPS $D3$-brane in the $NS5$-brane
ring background.
This model gives rise to a cos-type potential, which can lead to
inflation as the tachyon rolls down toward a potential minimum.
We carried out a careful analysis for the dynamics of inflation and
resulting density perturbations.

An important quantity which characterizes the amount of inflation
is the quantity $s$ defined in Eq.~(\ref{spara}).
We found that $s$ is constrained to be $s \gtrsim 30$ from
the observational data about the spectral index $n_{\rm S}$ of
scalar perturbations. The power spectrum approaches
a scale-invariant one for larger $s$ (see Fig.~\ref{nsfig}).
In the large $s$ limit the spectral index
takes a constant value given by Eq.~(\ref{limit})
($n_{\rm S}=0.978$ for the e-foldings $N=60$).
The tensor to scalar ratio grows for larger $s$, but there is
an upper limit $r_{{\rm max}}$ given by Eq.~(\ref{limit}).
Since $r_{{\rm max}}=0.044$ for $N=60$, this well
satisfies the recent observational constraint: $r<0.36$
at the $2\sigma$ level.

The amplitude of scalar perturbations also places constraints
on model parameters. We found that the number of $NS5$-branes
satisfies the condition (\ref{kcon1}) for any value of $s$
(see Fig.~\ref{kLMfig}).
{}From the requirement that the energy scale during
inflation does not exceed the Planck scale, we obtained an
upper limit of the number of $NS5$-branes as given in
Eq.~(\ref{knewcon}).
We note that the string coupling is weak ($g_{s} \ll 1$)
provided that $\sqrt{k}R/l_{s} \gg 10^4$.
Combining this with the throat condition $\sqrt{k}l_{s} \gg R$,
we found another constraint: $k \gg 10^4$.
This is well consistent with the condition (\ref{knewcon}).

In the vicinity of the ring of $NS5$-branes ($\rho \sim R$),
the tachyon potential is
approximately given as Eq.~(\ref{po3d}) by expanding
a Harmonic function around the ring.
This is connected to the potential inside and outside the
ring by imposing matching conditions.
We found that the energy scale at the potential minimum, $V_{1}$,
is $1/\sqrt{k}$ times smaller than the energy scale, $V_{0}$,
during inflation. If we use the constraint $k(l_s/l_p)^2 \lesssim
10^{27}$ coming from the condition for inflation, the
energy scale $V_{1}$ can not be very small to explain
dark energy. Although the tachyon exhibits oscillations
after inflation, the equation of state approaches $w_{\phi}=-1$
as illustrated in Fig.~\ref{phiw}.
The energy density of tachyon dominates
during radiation or matter dominant era, thus
disturbing the thermal history of the universe.
Hence we need to find a way to reduce the energy density $V_1$
to obtain a viable tachyon inflation scenario.

Although it is difficult to explain both inflation and dark energy
by using a single tachyon field, it is possible to make use of the
geometrical tachyon scenario for dark energy provided that
inflation is realized by another scalar field.
In this case we do not have the upper limit
given by Eq.~(\ref{knewcon}), since the geometrical tachyon
is not responsible for CMB anisotropies.
We derived the condition (\ref{dark}) for the number of $NS5$-branes
to realize the energy scale of dark energy observed today.
Although $k$ is required to be very large to satisfy this constraint,
it is interesting that the origin of dark energy can be explained
by using the geometrical tachyon.

While we have considered the motion of the BPS $D3$-brane in the
plane of the ring, it is possible to investigate the case in which
its motion is transverse to the ring.
Then the tachyon map yields a cosh-type potential \cite{TW2},
in which case we do not need to worry for the continuity
condition around the ring.
Since there remains an energy density $\tau_{3} R/\sqrt{kl_{s}^2}$
at the potential minimum ($\phi=0$), we expect that cosmological
evolution may  not change much compared to the model
discussed in this paper. Nevertheless it is certainly of interest
to investigate this case in more details to distinguish between
two geometrical tachyon scenarios.

\section*{ACKNOWLEDGMENTS}
We thank Steven Thomas and John Ward for useful
discussions. S.~T. is supported by JSPS
(Grant No.\,30318802).


\begin{thebibliography}{40}

\bibitem{senrev}
A.~Sen,
arXiv:hep-th/0410103.

\bibitem{DBI}
A.~Sen, JHEP {\bf 9910}, 008 (1999);
M.~R.~Garousi, Nucl. Phys. B{\bf 584}, 284 (2000);
Nucl. Phys. B {\bf 647}, 117 (2002);
JHEP {\bf 0305}, 058 (2003);
E.~A.~Bergshoeff, M.~de Roo, T.~C. de Wit,
E.~Eyras, S.~Panda, JHEP {\bf 0005}, 009 (2000);
J.~Kluson, Phys. Rev. D {\bf 62}, 126003 (2000);
D.~Kutasov and V.~Niarchos, Nucl. Phys. B {\bf 666},
56 (2003).

\bibitem{tachinfl}
A.~Mazumdar, S.~Panda and A.~Perez-Lorenzana,
Nucl.\ Phys.\ B {\bf 614}, 101 (2001); M.~Fairbairn and
M.~H.~G.~Tytgat,
Phys.\ Lett.\ B {\bf 546}, 1 (2002); A.~Feinstein,
Phys.\ Rev.\ D {\bf 66}, 063511 (2002); M.~Sami, P.~Chingangbam
and T.~Qureshi,
Phys.\ Rev.\ D {\bf 66}, 043530 (2002); M.~Sami,
Mod.\ Phys.\ Lett.\ A {\bf 18}, 691 (2003); Y.~S.~Piao,
R.~G.~Cai, X.~m.~Zhang and Y.~Z.~Zhang,
Phys.\ Rev.\ D {\bf 66}, 121301 (2002).

\bibitem{KL}
L.~Kofman and A.~Linde,
JHEP {\bf 0207}, 004 (2002).

\bibitem{tachyonpapers}
G.~W.~Gibbons,
Phys.\ Lett.\ B {\bf 537}, 1 (2002);
S.~Mukohyama,
Phys.\ Rev.\ D {\bf 66}, 024009 (2002);
Phys.\ Rev.\ D {\bf 66}, 123512 (2002);
D.~Choudhury, D.~Ghoshal, D.~P.~Jatkar and S.~Panda,
Phys.\ Lett.\ B {\bf 544}, 231 (2002);
G.~Shiu and I.~Wasserman,
Phys.\ Lett.\ B {\bf 541}, 6 (2002);
T.~Padmanabhan,
Phys.\ Rev.\ D {\bf 66}, 021301 (2002);
J.~S.~Bagla, H.~K.~Jassal and T.~Padmanabhan,
Phys.\ Rev.\ D {\bf 67}, 063504 (2003);
G.~N.~Felder, L.~Kofman and A.~Starobinsky,
JHEP {\bf 0209}, 026 (2002);
J.~M.~Cline, H.~Firouzjahi and P.~Martineau,
JHEP {\bf 0211}, 041 (2002);
M.~C.~Bento, O.~Bertolami and A.~A.~Sen,
Phys.\ Rev.\ D {\bf 67}, 063511 (2003);
J.~g.~Hao and X.~z.~Li,
Phys.\ Rev.\ D {\bf 66}, 087301 (2002);
C.~j.~Kim, H.~B.~Kim and Y.~b.~Kim,
Phys.\ Lett.\ B {\bf 552}, 111 (2003);
T.~Matsuda,
Phys.\ Rev.\ D {\bf 67}, 083519 (2003);
A.~Das and A.~DeBenedictis,
arXiv:gr-qc/0304017;
Z.~K.~Guo, Y.~S.~Piao, R.~G.~Cai and Y.~Z.~Zhang,
Phys.\ Rev.\ D {\bf 68}, 043508 (2003);
L.~R.~W.~Abramo and F.~Finelli,
Phys.\ Lett.\ B {\bf 575} (2003) 165;
G.~W.~Gibbons,
Class.\ Quant.\ Grav.\  {\bf 20}, S321 (2003);
M.~Majumdar and A.~C.~Davis,
arXiv:hep-th/0304226;
S.~Nojiri and S.~D.~Odintsov,
Phys.\ Lett.\ B {\bf 571}, 1 (2003);
E.~Elizalde, J.~E.~Lidsey, S.~Nojiri and S.~D.~Odintsov,
Phys.\ Lett.\ B {\bf 574}, 1 (2003);
D.~A.~Steer and F.~Vernizzi,
Phys.\ Rev.\ D {\bf 70}, 043527 (2004);
V.~Gorini, A.~Y.~Kamenshchik, U.~Moschella and V.~Pasquier,
Phys.\ Rev.\ D {\bf 69}, 123512 (2004);
L.~P.~Chimento,
Phys.\ Rev.\ D {\bf 69}, 123517 (2004);
J.~M.~Aguirregabiria and R.~Lazkoz,
Phys.\ Rev.\ D {\bf 69}, 123502 (2004); M.~B.~Causse,
arXiv:astro-ph/0312206;
B.~C.~Paul and M.~Sami,
Phys.\ Rev.\ D {\bf 70}, 027301 (2004);
G.~N.~Felder and L.~Kofman,
Phys.\ Rev.\ D {\bf 70}, 046004 (2004);
J.~M.~Aguirregabiria and R.~Lazkoz,
Mod.\ Phys.\ Lett.\ A {\bf 19}, 927 (2004);
L.~R.~Abramo, F.~Finelli and T.~S.~Pereira,
Phys.\ Rev.\ D {\bf 70}, 063517 (2004);
G.~Calcagni,
Phys.\ Rev.\ D {\bf 70}, 103525 (2004);
G.~Calcagni and S.~Tsujikawa,
Phys.\ Rev.\ D {\bf 70}, 103514 (2004); P.~F.~Gonzalez-Diaz,
Phys.\ Rev.\ D {\bf 70}, 063530 (2004);
S.~K.~Srivastava,
arXiv:gr-qc/0409074; gr-qc/0411088;
P.~Chingangbam and T.~Qureshi,
arXiv:hep-th/0409015;
S.~Tsujikawa and M.~Sami,
Phys.\ Lett.\ B {\bf 603}, 113 (2004); M.~R.~Garousi, M.~Sami and
S.~Tsujikawa, Phys.\ Lett.\ B {\bf 606}, 1 (2005); Phys.\ Rev.\ D
{\bf 71}, 083005 (2005); N.~Barnaby and J.~M.~Cline,
Int.\ J.\ Mod.\ Phys.\ A {\bf 19}, 5455 (2004);
E.~J.~Copeland, M.~R.~Garousi, M.~Sami and S.~Tsujikawa,
Phys.\ Rev.\ D {\bf 71}, 043003 (2005);
B.~Gumjudpai, T.~Naskar, M.~Sami and S.~Tsujikawa,
JCAP {\bf 0506}, 007 (2005); I. P. Neupane, CQG {\bf 21}, 4383 (2004);
M.~Novello, M.~Makler, L.~S.~Werneck and C.~A.~Romero,
Phys.\ Rev.\ D {\bf 71}, 043515 (2005);
A.~Das, S.~Gupta, T.~D.~Saini and S.~Kar,
Phys.\ Rev.\ D {\bf 72}, 043528 (2005);
H.~Singh,
arXiv:hep-th/0505012;
S.~Tsujikawa,
arXiv:astro-ph/0508542.


\bibitem{BA}
C. P. Burgess, M. Majumdar, D. Nolte, F. Quevedo, G. Rajesh and
R. J. Zhang, JHEP {\bf 0107} (2001) 047; C. P. Burgess, P.
Martineau, F. Quevedo, G. Rajesh and R. J. Zhang, JHEP {\bf 0203}
(2002) 052; D. Choudhury, D. Ghoshal, D. P. Jatkar and S. Panda,
JCAP {\bf 0307} (2003) 009.

\bibitem{GST}
M.~R.~Garousi, M.~Sami and S.~Tsujikawa,
Phys.\ Rev.\ D {\bf 70}, 043536 (2004).

\bibitem{warpcosm}
J. Raemaekers, JHEP {\bf 0410}, 057 (2004); P. Chingangbam, A.
Deshamukhya and S. Panda, JHEP {\bf 0502}, 052 (2005); D.
Cremades, F. Quevedo and A. Sinha, arXiv:hep-th/0505252.

\bibitem{sen}
A.~Sen, JHEP {\bf 0204}, 048 (2002); JHEP
{\bf 0207}, 065 (2002);
Mod. Phys. Lett. A {\bf 17}, 1797 (2002);
arXiv: hep-th/0312153.

\bibitem{kutasov}
D. Kutasov, arXiv:hep-th/0405058, arXiv:hep-th/0408073.

\bibitem{TW1}
S.~Thomas and J.~Ward,
JHEP {\bf 0502}, 015 (2005).

\bibitem{TW2}
S.~Thomas and J.~Ward,
arXiv:hep-th/0502228.

\bibitem{NS5cos}
H. Yavartano, arXiv:hep-th/0407079; A. Ghodsi and A. E. Mosaffa,
Nucl. Phys. B {\bf 714} (2005) 30.

\bibitem{TWinf}
S.~Thomas and J.~Ward,
arXiv:hep-th/0504226.

\bibitem{Callan}
C.~G.~.~Callan, J.~A.~Harvey and A.~Strominger,
Nucl.\ Phys.\ B {\bf 367}, 60 (1991);
hep-th/9112030.

\bibitem{sfet}
K. Sfetsos, arXiv:hep-th/9903201.

\bibitem{MFB}
V.~F.~Mukhanov, H.~A.~Feldman and R.~H.~Brandenberger,
Phys.\ Rept.\  {\bf 215}, 203 (1992);
H.~Kodama and M.~Sasaki,
Prog.\ Theor.\ Phys.\ Suppl.\  {\bf 78}, 1 (1984);
B.~A.~Bassett, S.~Tsujikawa and D.~Wands,
arXiv:astro-ph/0507632.

\bibitem{HN}
J.~c.~Hwang and H.~Noh,
Phys.\ Rev.\ D {\bf 66}, 084009 (2002).

\bibitem{SV}
D.~A.~Steer and F.~Vernizzi,
Phys.\ Rev.\ D {\bf 70}, 043527 (2004).

\bibitem{obcon}
H.~V.~Peiris \textit{et al.}, Astrophys. J. Suppl.
\textbf{148}, 213 (2003);
V.~Barger, H.~S.~Lee, and D.~Marfatia,
Phys.\ Lett. \ B \textbf{565}, 33 (2003);
W.~H.~Kinney, E.~W.~Kolb, A.~Melchiorri and A.~Riotto,
Phys.\ Rev.\ D \textbf{69}, 103516 (2004);
S.~M.~Leach and A.~R.~Liddle, Phys. Rev. D
\textbf{68}, 123508 (2003);
M. Tegmark {\em et al.},
Phys.\ Rev.\ D {\bf 69}, 103501 (2004);
S.~Tsujikawa and A.~R.~Liddle,
JCAP {\bf 0403}, 001 (2004);
S.~Tsujikawa and B.~Gumjudpai,
Phys.\ Rev.\ D {\bf 69}, 123523 (2004).

\bibitem{obcon2}
U.~Seljak {\it et al.},
Phys.\ Rev.\ D {\bf 71}, 103515 (2005).

\bibitem{FKS}
A.~V.~Frolov, L.~Kofman and A.~A.~Starobinsky,
Phys.\ Lett.\ B {\bf 545}, 8 (2002).


\end{thebibliography}
\end{document}